\documentclass[10pt, conference, letterpaper]{IEEEtran} 
\usepackage{dblfloatfix}    
\IEEEoverridecommandlockouts
\usepackage{cite}
\usepackage{amsmath,amssymb,amsfonts}
\usepackage{algorithmic}
\usepackage{graphicx}
\usepackage{textcomp}
\usepackage{xcolor}
\usepackage{multirow}
\usepackage{array}
\usepackage{longtable}
\usepackage{booktabs}
\usepackage{subfigure} 
\usepackage{url} 
\usepackage[letterpaper, right=0.62in, left=0.62in, top=0.7in, bottom=0.95in]{geometry}
\def\BibTeX{{\rm B\kern-.05em{\sc i\kern-.025em b}\kern-.08em
    T\kern-.1667em\lower.7ex\hbox{E}\kern-.125emX}}
\begin{document}

\title{Digital Twins for Autonomous Driving: A Comprehensive Implementation and Demonstration\\
}

\author{
    \IEEEauthorblockN{Kui Wang\IEEEauthorrefmark{1}, Tao Yu\IEEEauthorrefmark{1}, Zongdian Li\IEEEauthorrefmark{1}, Kei Sakaguchi\IEEEauthorrefmark{1}, Omar Hashash\IEEEauthorrefmark{2}, and Walid Saad\IEEEauthorrefmark{2}}
    
    \IEEEauthorblockA{\IEEEauthorrefmark{1}Department of Electrical and Electronic Engineering, Tokyo Institute of Technology, Tokyo, Japan.}
    
    \IEEEauthorblockA{\IEEEauthorrefmark{2}Wireless@VT, Bradley Department of Electrical and Computer Engineering, Virginia Tech, Arlington, VA, USA.}
    \IEEEauthorblockA{Emails: \IEEEauthorrefmark{1}\{kuiw, yutao, lizd, sakaguchi\}@mobile.ee.titech.ac.jp, \IEEEauthorrefmark{2}\{omarnh, walids\}@vt.edu}}

\maketitle

\begin{abstract}
The concept of a digital twin (DT) plays a pivotal role in the ongoing digital transformation and has achieved significant strides for various wireless applications in recent years. In particular, the field of autonomous vehicles is a domain that is ripe for exploiting the concept of DT. Nevertheless, there are many challenges that include holistic consideration and integration of hardware, software, communication methods, and collaboration of edge/cloud computing. In this paper, an end-to-end (E2E) real-world smart mobility DT is designed and implemented for the purpose of autonomous driving. The proposed system utilizes roadside units (RSUs) and edge computing to capture real-world traffic information, which is then processed in the cloud to create a DT model. This DT model is then exploited to enable route planning services for the autonomous vehicle to avoid heavy traffic. Real-world experimental results show that the system reliability can reach $99.53\%$ while achieving a latency that is $3.36\%$ below the 3GPP recommended value of $100$~ms for autonomous driving. These results clearly validate the effectiveness of the system according to practical 3GPP standards for sensor and state map sharing (SSMS) and information sharing.

\end{abstract}


\section{Introduction}
The realm of the Internet of Things (IoT) is rapidly expanding beyond the conventional frameworks, branching out to encapsulate a broader spectrum of digital connectivity \cite{b2}. This emerging paradigm facilitates intelligent data sharing among entities, thereby advancing our ability to monitor, control, and optimize the physical world. As a prominent manifestation of IoT, digital twin (DT) technology has begun to establish its foothold across different industries \cite{b3}. The potential of DTs lies in their capabilities to construct comprehensive digital representations, thereby enabling bidirectional interaction between the physical space and cyber space \cite{add}. Henceforth, DTs empower real-time decision-making and enhance efficiency, productivity, and adaptability in a multitude of applications \cite{b4}.

With the rapid advancement in autonomous driving technologies, the convergence of DTs and autonomous vehicles presents an exciting new frontier. The implementation of DTs within the autonomous driving ecosystem creates vast possibilities for improved safety, efficiency, and robustness of the system \cite{b5}. As such, leveraging real-time synchronization between the digital and physical worlds, autonomous vehicles can access global information to "see more and see further", enhancing their awareness and understanding of the traffic environment \cite{b1}. Furthermore, a DT allows for effective testing and validation of autonomous driving algorithms in a controllable and scalable virtual environment, accelerating the development and deployment of autonomous driving \cite{b6}.

Hence, there has been a growing interest in introducing the concept of DTs in smart mobility \cite{b7,b8,b9,b10,b11,b12,b13,b14,b15}. The mainstream roles of vehicular DTs are discussed and summarized in \cite{b7}. The authors in \cite{b8} proposed a DT-enabled scheduling architecture to help multiple vehicle users fulfill their personalized requirements for path planning. Nevertheless, this prior work \cite{b8} was only conducted and evaluated based on simulation results, without realizing path planning for vehicles in a real-world setting. Recent works in \cite{b9,b10,b11} investigated the generation of highly accurate DT models by using various onboard and infrastructure sensors with edge computing to collect and process real-time information from the physical space. However, the decision feedback from cyber space to entities in physical space is not properly captured. In \cite{b12,b13,b14}, the implementation of DT systems was done in order to provide situational awareness and cooperative driving services for \emph{human} drivers, which fails to consider the \emph{autonomy} of vehicles. To the best of our knowledge, there is a lack of the holistic integration of \emph{DTs with autonomous vehicles}, as well as an absence of any practical end-to-end (E2E) system designs and implementation of such a system. Hence, in our previous work \cite{b15}, we proposed a system architecture for the smart mobility DT and conducted a preliminary proof-of-concept (PoC) test. Our PoC provides a route planning service that allows a vehicle to avoid overcrowded traffic. However, this fundamental work in \cite{b15} only considered offline collection of sensor data and overlooked the route planning decisions and actions that were to be executed at the vehicle level.

In contrast to these prior works, the main contribution of this paper is a real-time smart mobility DT framework for autonomous driving, as well as a comprehensive real-world implementation of this system. The proposed system requires a realistic virtual representation of real-life traffic in autonomous driving scenarios as well as the establishment of an effective and stable request-feedback loop for the autonomous vehicle to obtain cloud-based decisions. Thus, from a system design perspective, the challenge lies in holistic consideration and integration of hardware, software, communication methods, and collaboration of edge/cloud computing. In addition, it is important to implement and test the whole system in real-time traffic conditions, which allows us to evaluate and validate the performance of the designed DT. Hence, the system testing is carried out in different scenarios to provide a thorough analysis on the impact of various factors. In summary, our key contributions include:

\begin{figure}[t]
    \centerline{\includegraphics[width=0.48\textwidth]{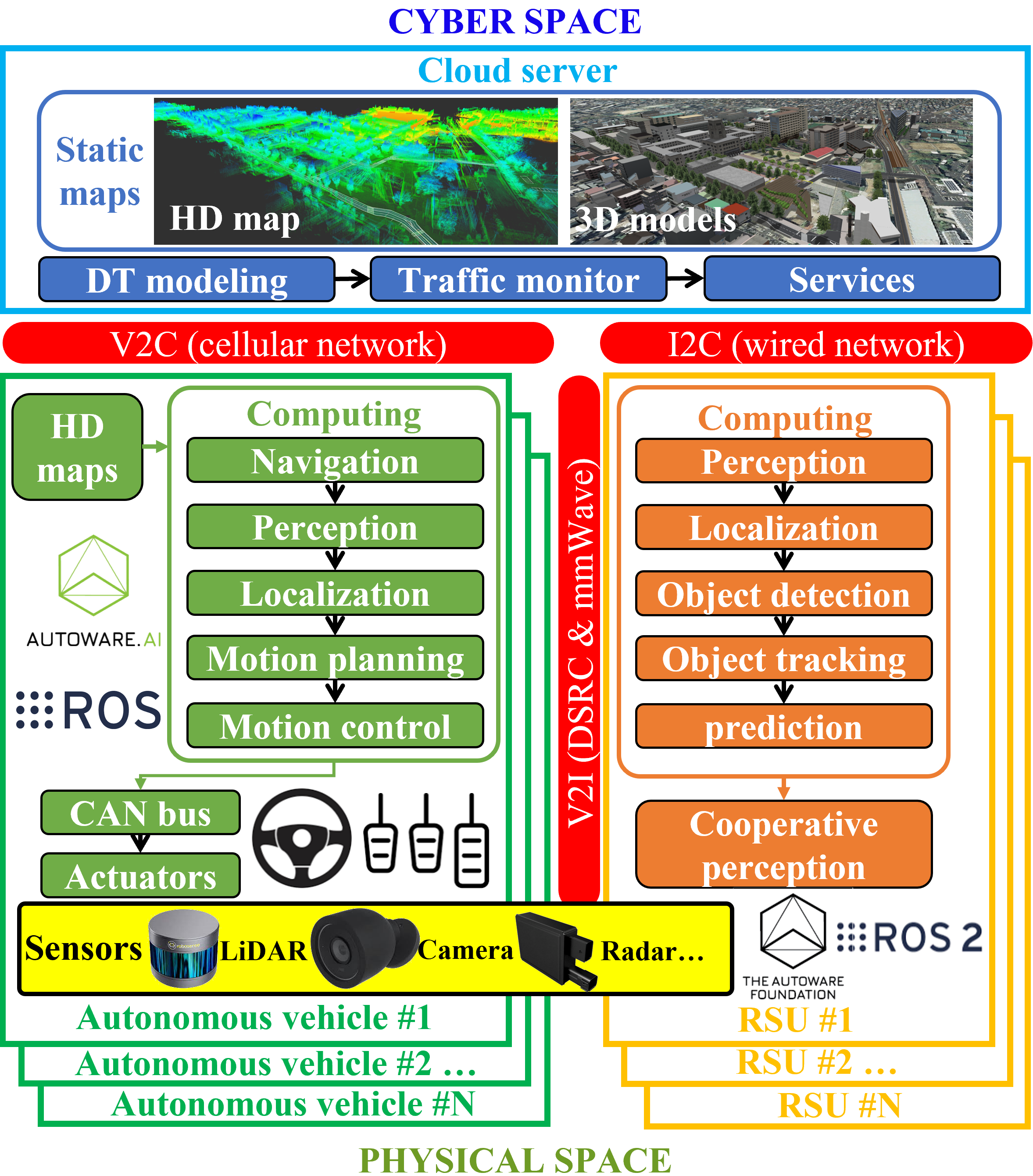}}
    \caption{System architecture of smart mobility DT.}
    \label{fig: dt-design}
\end{figure}

\begin{itemize}
    \item We design a smart mobility DT that reflects objects of interest in the context of road driving scenarios, e.g., roads, infrastructures, vehicles, and pedestrians.
    \item We implement the DT platform for autonomous driving by employing the roadside units (RSUs) and the autonomous vehicle equipped with sensors, communication modules, and edge computing capabilities. 
    \item We design a cloud-based route planning service for the autonomous vehicle based on real-time traffic information.
    \item Our results show the realization of the DT modeling and demonstrate the effectiveness of the route planning service, whereby the planned routes are successfully transmitted to the vehicle for real-time execution.
    \item We validate the system performance in terms of reliability and latency based on 3GPP standards for sensor and state map sharing (SSMS) and information sharing.
\end{itemize}

The rest of the paper is organized as follows. Section II provides a concise overview of the system design. Details about hardware deployment, software installation, and route planning are shown in Section III. Section IV discusses the experimental evaluation. Section V concludes the paper.

\section{Design for Smart Mobility DT}

We first develop and propose our DT system architecture that is tailored for implementation in a specific but generalizable environment. The proposed smart mobility DT system is essentially based on our work in \cite{b15, yu2023internet}. This proposed system enables real-time DT modeling and real-time feedback services to the autonomous vehicle.

Fig.~\ref{fig: dt-design} gives the system architecture, consisting of autonomous vehicles and RSUs in the physical world and a DT system maintained within the cloud server. We exploit the computing capabilities of the edge and cloud by distributing various functions and services on these two planes. The edge, referring to RSUs and autonomous vehicles, provides better real-time operation in order to address delay-sensitive tasks such as environmental perception and vehicle maneuvering. Meanwhile, the cloud offers robust data processing capabilities with higher latency, making it suitable for handling large-scale and computation-intensive tasks that need global information, such as DT modeling and deriving cloud-based services. Thus, our system design incorporates edge and cloud computing to enhance efficiency and safety for autonomous driving.

In the physical world, diverse sensors on the vehicle and RSUs capture real-time data from their surroundings. Then the vehicle can achieve a high level of autonomy, incorporating functional modules such as navigation, perception, localization, motion planning, and control. Similarly, RSUs use sensors for object detection, tracking, and prediction, upload the perception data to the cloud, and share it with nearby vehicles to enhance their safety during the driving process.

The smart mobility DT within the cloud plane is dynamically updated using real-time perception data from the physical system, thereby accurately mirroring the current state of the real-world traffic. After receiving the real-time perception data from RSU edges, the cloud server synchronizes incoming edge
channels and locates detected objects based on their relative coordinates to different edge sensors. By integrating with static HD maps and 3D models, we can realize the generation of a virtual representation from the physical space to cyber space. By utilizing the smart mobility DT, we can facilitate the provision of a variety of derivative services for autonomous vehicles that include route planning, risk prediction, and congestion alerting. Since these services are predicated on the comprehensive
global information from the DT, they can assist vehicles in evading high-risk and congested areas outside onboard sensing horizons and improving their overall commuting efficiency.

It is important to note that the communication within this system is based on a heterogeneous vehicle-to-everything (V2X) network \cite{li1012hetsdvn}, which includes vehicle-to-infrastructure (V2I), vehicle-to-cloud (V2C), and infrastructure-to-cloud (I2C) communications. Considering the large communication distance between vehicles and the cloud, the cellular network is applied for V2C communication. Depending on the application scenario, RSUs may send data of different levels to the cloud server and nearby vehicles, such as raw data or processed data, which have distinct requirements on communication coverage and bandwidth. Hence, we use wired networks for I2C communication and employ both dedicated short-range communications (DSRC) and millimeter wave (mmWave) technologies for V2I. 

\begin{table*}[t]
\centering
\caption{Communication requirements and performances}
\label{tab:my-table}
\begin{tabular}{c|cccc}
\hline
Function             & Requirements                                                                                      & Message contents                                                                                                                       & Methods      & Performances                                                                       \\ \hline
V2C                  & \begin{tabular}[c]{@{}c@{}}Speed: $\geq 80$~Mbps\\ Coverage: $\geq 1$~km\end{tabular} & \begin{tabular}[c]{@{}c@{}}Upload: vehicle position and motion states. \\ Download: cloud-based decisions\end{tabular}        & WiMAX        & \begin{tabular}[c]{@{}c@{}}$\geq 120$~Mbps\\ $\geq 50$~km\end{tabular} \\ \hline
\multirow{2}{*}{V2I} & \begin{tabular}[c]{@{}c@{}}Speed: $\geq 1$~Mbps\\ Coverage: $\geq 50$~m\end{tabular}  & \begin{tabular}[c]{@{}c@{}}Cooperative perception (processed data, \\ e.g., detection and tracking result)\end{tabular}       & Wi-Fi router & \begin{tabular}[c]{@{}c@{}}$\geq 10$~Mbps\\ $\geq 200$~m\end{tabular}   \\ \cline{2-5} 
                     & \begin{tabular}[c]{@{}c@{}}Speed: $\geq 1$~Gbps\\ Coverage: $\geq 20$~m\end{tabular}  & \begin{tabular}[c]{@{}c@{}}Cooperative perception (raw sensor data, e.g. \\ LiDAR point cloud and camera images)\end{tabular} & WiGig        & \begin{tabular}[c]{@{}c@{}}$\geq 1$~Gbps\\ $\geq 120$~m\end{tabular}    \\ \hline
I2C                  & Speed: $\geq 1$~Gbps                                                                        & Environmental perception (both raw data and processed data)                                                                   & Ethernet     & $\geq 1$~Gbps                                                                \\ \hline
\end{tabular}
\end{table*}

\section{Implementation for Smart Mobility DT}

In this section, we will discuss the implementation of our system in the Tokyo Tech smart mobility field, as well as the route planning workflow involved.

\subsection{Hardware Deployment}

\begin{figure}[t]
    \centerline{\includegraphics[width=0.49\textwidth]{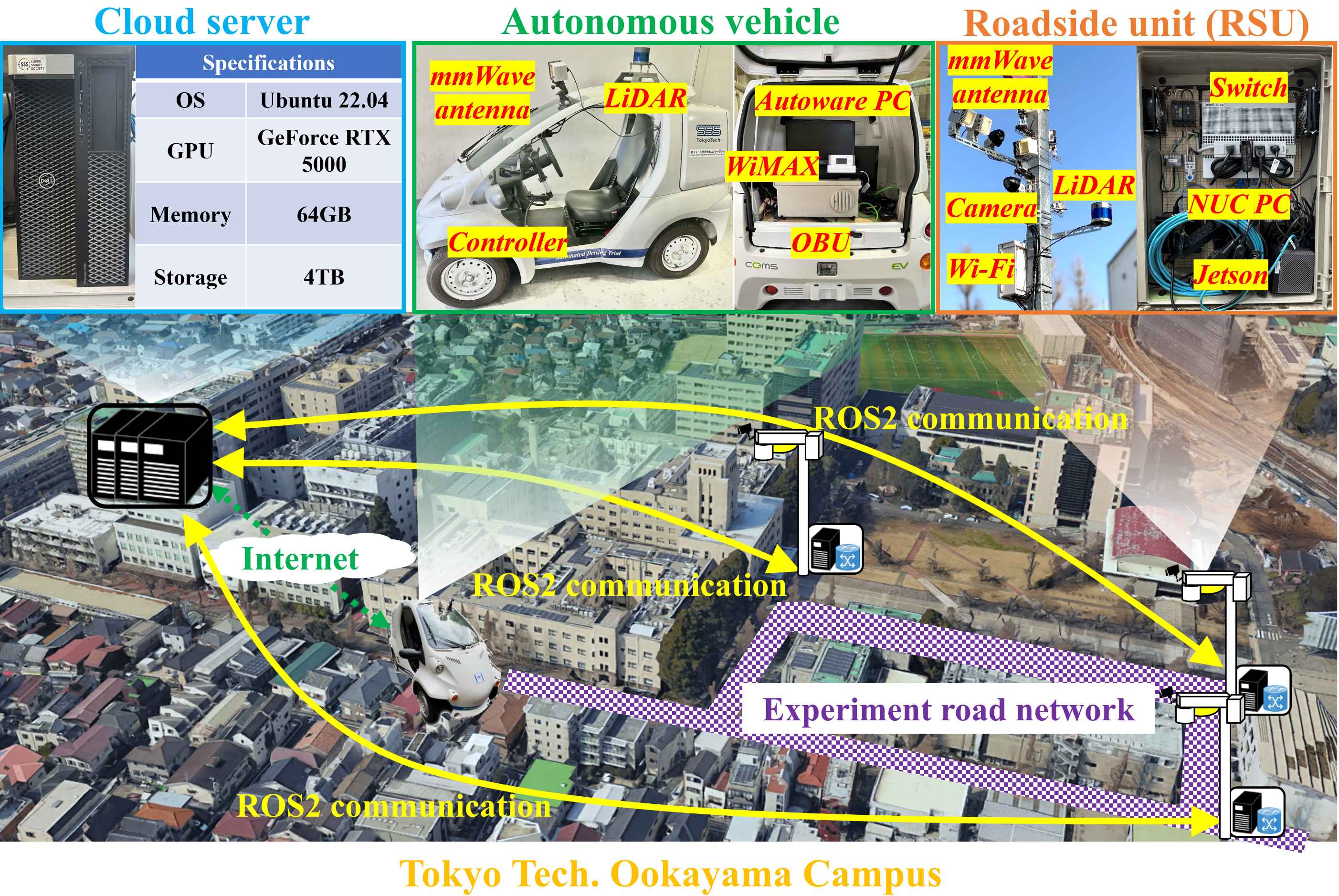}}
    \caption{Hardware components and Tokyo Tech. smart mobility field.}
    \label{fig: implementation}
\end{figure}

Our testing field in Tokyo Tech Ookayama campus, as well as our hardware devices, are shown in Fig.~\ref{fig: implementation}. In the field setting, we have three RSUs and one autonomous vehicle. The three RSUs are located at the corners of a square road section within the campus. Each RSU is equipped with an 80-layer LiDAR. These sensors collect raw data from the physical world, capturing information about dynamic objects in the traffic environment. Additionally, each RSU is also outfitted with edge computing devices, i.e., NVIDIA Jetson AGX Orin, where we launch functional modules for functions like object detection, tracking, and prediction, among others. The communication between RSUs and the cloud server relies on the campus Ethernet network. To cater to the requirements of different application scenarios, we upload varying levels of perception data, such as raw point cloud data, detection and tracking results to cloud. 

The autonomous vehicle is equipped with a 32-layer LiDAR sensor positioned on the rooftop to sense its surroundings. A dedicated Autoware PC is utilized for processing the LiDAR data and performing environmental perception, localization, motion planning, and motion control. The control signals are then transmitted to the vehicle's onboard unit (OBU) to enable autonomous driving. Additionally, the vehicle is equipped with communication modules, including worldwide interoperability for microwave access (WiMAX) for V2C communication, a Wi-Fi router as a replacement for dedicated short-range communications (DSRC), and a WiGig antenna for millimeter wave (mmWave) communication. The selection of communication equipment is based on specific requirements and functions. Table I shows the required communication speeds and coverage in our experiment field, as well as the performance of selected communication methods. These devices effectively facilitate information sharing among edge and cloud planes.

The cloud server works as the central hub for data aggregation and storage, global information processing, and providing feedback services to the autonomous vehicle. Therefore, the performance requirements for the cloud server are very high, which encompass scalability, processing power, and reliability to efficiently and securely handle large volumes of real-time data. Hence, we opted for a computer equipped with a GeForce RTX 5000 GPU, ensuring ample memory and storage capacity to meet the demanding computational requirements. 

\subsection{Software Installation}

The software in our platform is centered around two key open-source software, i.e., Autoware \cite{Autoware} and Robot Operating System (ROS).

In the RSU edges and cloud server, we deploy Autoware Universe and version 2 of the ROS (ROS2). Autoware Universe serves as the primary software framework for our system. It provides comprehensive functionality for the development of autonomous driving systems, making it well-suited for DT modeling. Autoware facilitates the detection and tracking of traffic participants within the RSU sensor range. The detection module applies the CenterPoint framework\cite{b17}, which can detect, identify, and visualize 3D objects from the LiDAR point clouds in real-time. Then, a tracking algorithm, named Multi-object Tracker \cite{b19}, is responsible for assigning the detected objects with IDs and estimating their velocities. Moreover, Autoware Universe is built on ROS2, which enables easy deployment and seamless communication among distributed computers. In our case, both the cloud server and the RSU edge computing devices are connected to the campus Ethernet network and assigned IP addresses within a local area network (LAN). Hence, we can subscribe to all the ROS2 topics on the cloud server. The utilization of ROS2 greatly facilitates the modeling of the DT and further enhances the interoperability and scalability of the DT system. After acquiring object tracking data from three RSUs, the cloud server utilizes objects' positions to align them with the corresponding road segments, which facilitates traffic monitoring and the computation of congestion level on each road section.

The Autoware PC onboard the autonomous vehicle is installed with an Ubuntu 16.04 operating system, enabling the implementation of autonomous driving functionalities using Autoware AI based on ROS. Autoware AI is responsible for driving the vehicle's LiDAR sensor and utilizes the normal distributions transform (NDT) algorithm for LiDAR scan matching with the 3D point cloud map, which enables real-time localization with centimeter-level accuracy. The route from origin to destination is in the form of a prerecorded waypoint file, consisted of a set of route points with position, velocity, and orientation information. The automated vehicle can sequentially track these waypoints to follow any given route. To perform local motion planning and motion control, a velocity planner adjusts the velocity plan based on the waypoints to decelerate or accelerate in response to nearby objects and road characteristics, including stop lines and traffic lights. Then, a pure pursuit algorithm is employed to generate coordinated sets of velocities and steering angles that guide the ego vehicle to follow the target waypoints.

\subsection{Route Planning Scenario and Workflow}

\begin{figure}[t]
    \centerline{\includegraphics[width=0.48\textwidth]{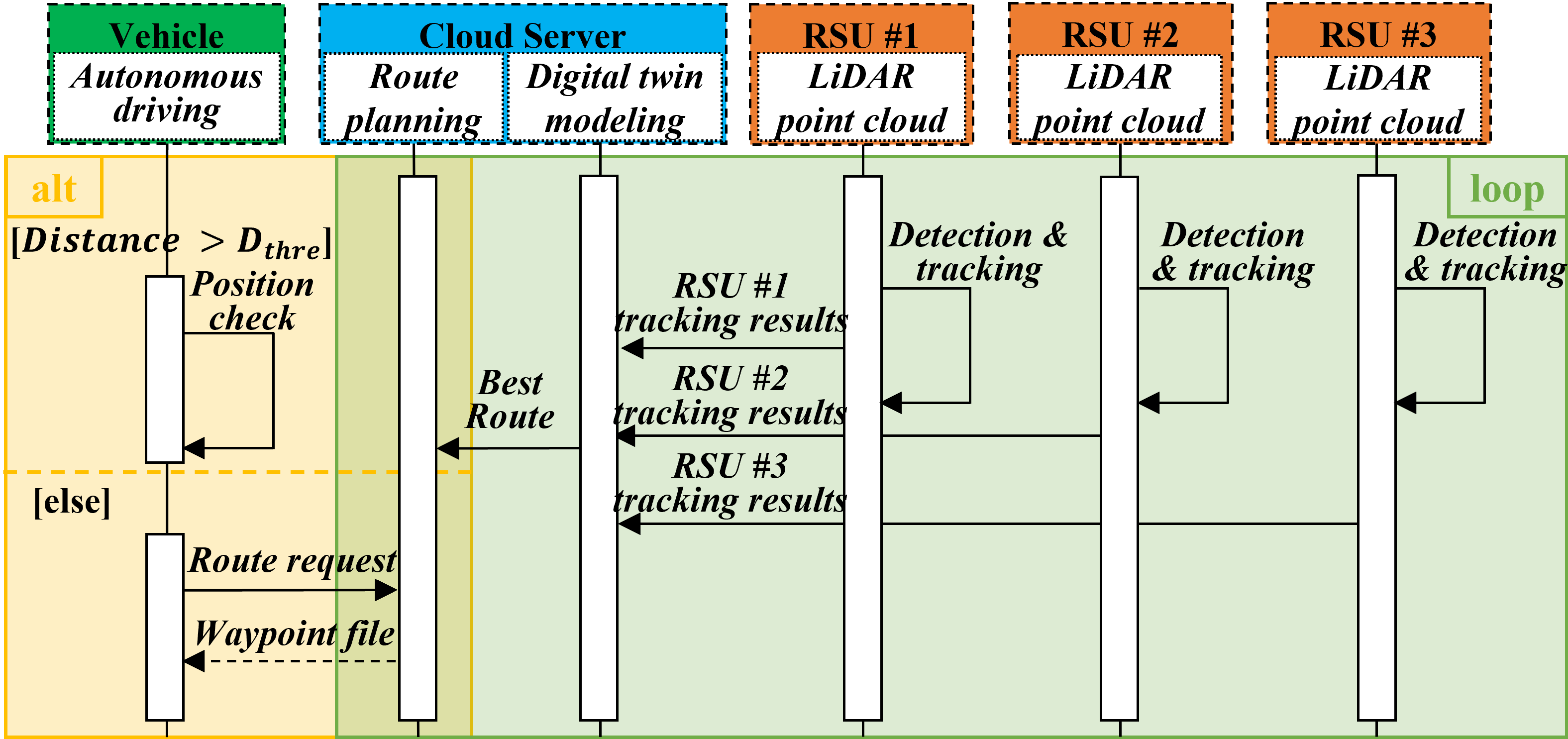}}
    \caption{Sequence diagram of route planning service.}
    \label{fig: workflow}
\end{figure}

Based on the hardware and software implementation, we design the workflow of the route planning service, as shown in Fig.~\ref{fig: workflow}. There are two main sequences in the workflow, i.e., DT modeling and route requisition. The DT modeling sequence is a loop fragment for real-time edge sensing, detection, tracking, and data acquisition on the cloud plane. Three RSUs run the LiDAR driver, object detection, and tracking. The detection and tracking results are transmitted from the RSUs to the cloud. Then the cloud computing will help the ego vehicle choose the optimal path to circumvent traffic congestion and to improve its efficiency, using the shortest-paths algorithm like Dijkstra's algorithm and A* algorithm. Since the specific route planning is not the primary focus of this paper, it is not elaborated upon in detail here.

The route requisition sequence is an alternative combination fragment, obeying the “if then else” logic. During the vehicle's movement from the starting point to the destination, it utilizes LiDAR-based localization and the road vector map to calculate the distance between itself and upcoming intersections. When the vehicle approaches an intersection, within a predetermined threshold $D_{\textrm{thre}}$, it initiates a path request to the cloud server via the hypertext transfer protocol (HTTP). The HTTP server running on the cloud receives the request and sends the uniform resource locator (URL) of the optimal path to the vehicle. Then the autonomous vehicle downloads the route and tracks it for vehicle navigation.

\section{Experimental Demonstration}

In this section, we will show the outdoor test results and the evaluation from the aspects of functionality and reliability.

\subsection{Field Experiment Results}

\begin{figure}[t]
\centering
\subfigure[]{
\begin{minipage}[b]{0.47\textwidth}
\includegraphics[width=1\textwidth]{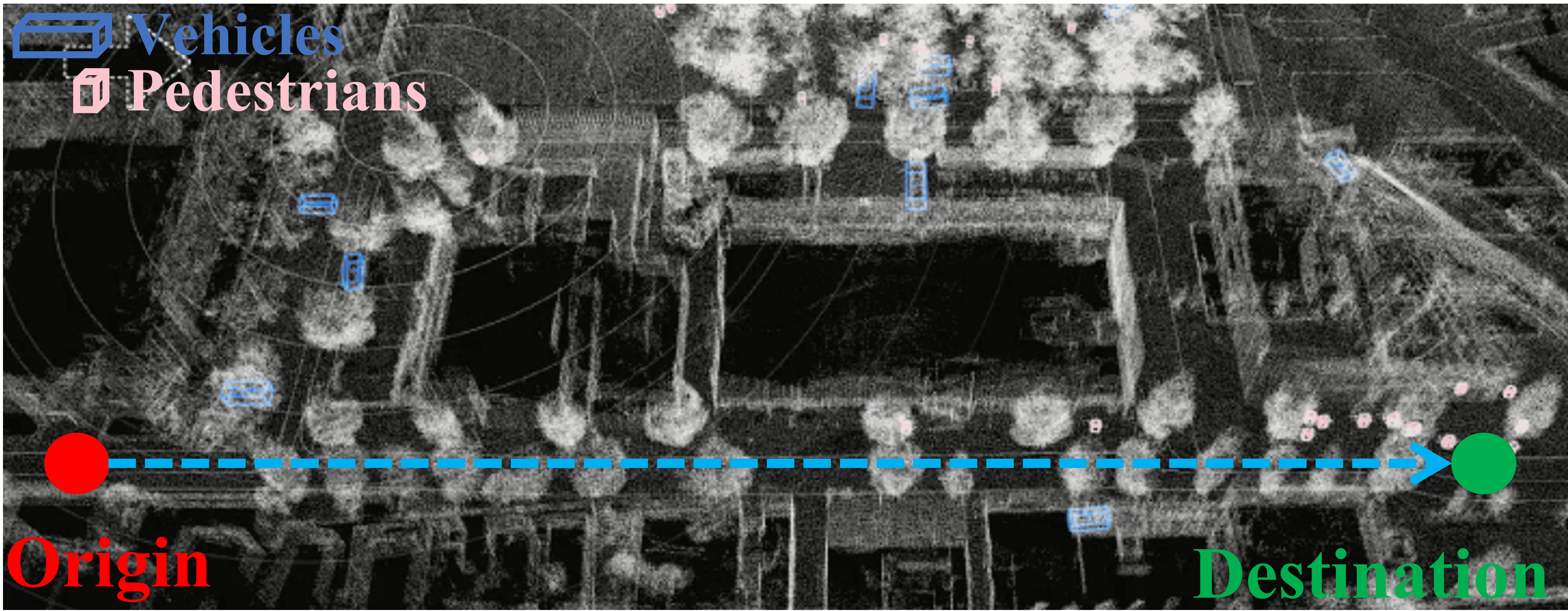} 
\end{minipage}
}
\subfigure[]{
\begin{minipage}[b]{0.47\textwidth}
\includegraphics[width=1\textwidth]{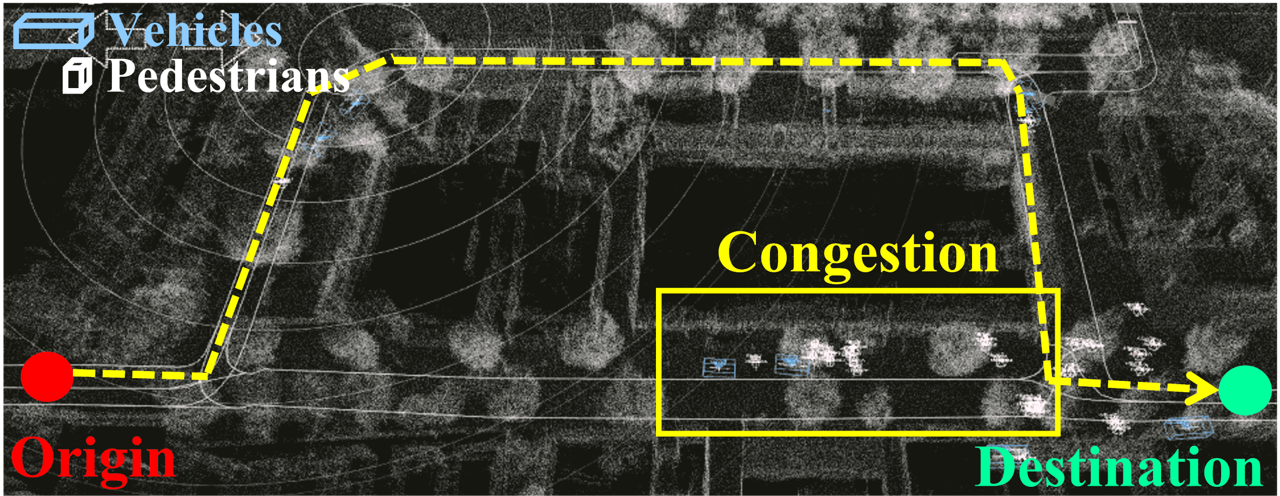} 
\end{minipage}
}
\caption{Real-time DT modeling: (a) Scenario \#A: no congestion in the road network, (b) Scenario \#B: congestion occurs on the straight route}
\label{fig: cloudcases}
\end{figure}

\begin{figure*}[t]
\centering
\subfigure[]{
\begin{minipage}[b]{0.98\textwidth}
\includegraphics[width=1\textwidth]{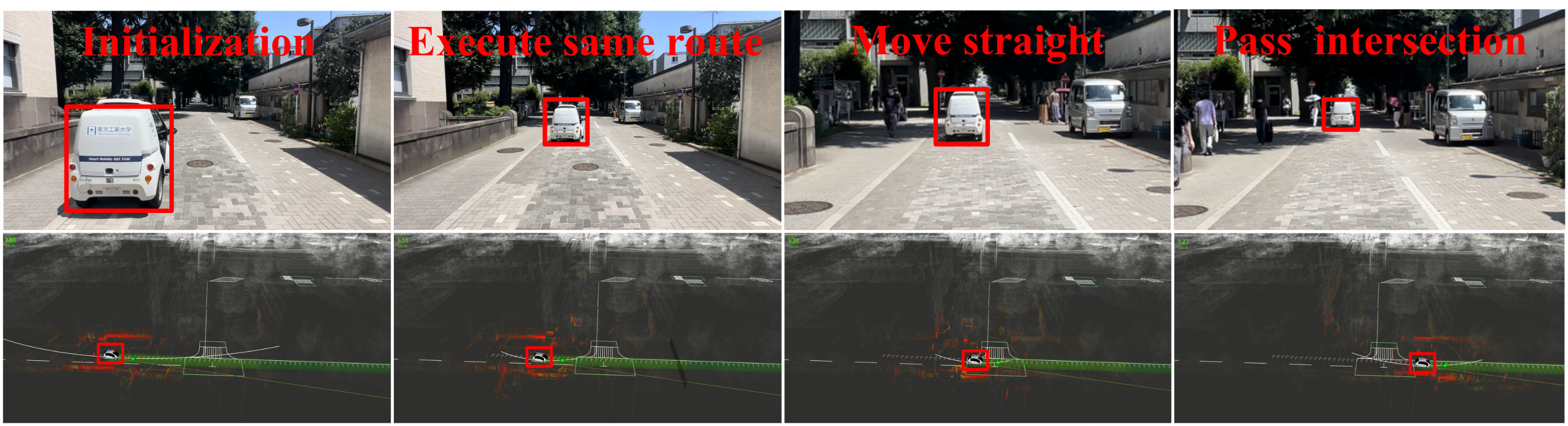} 
\end{minipage}
}
\subfigure[]{
\begin{minipage}[b]{0.98\textwidth}
\includegraphics[width=1\textwidth]{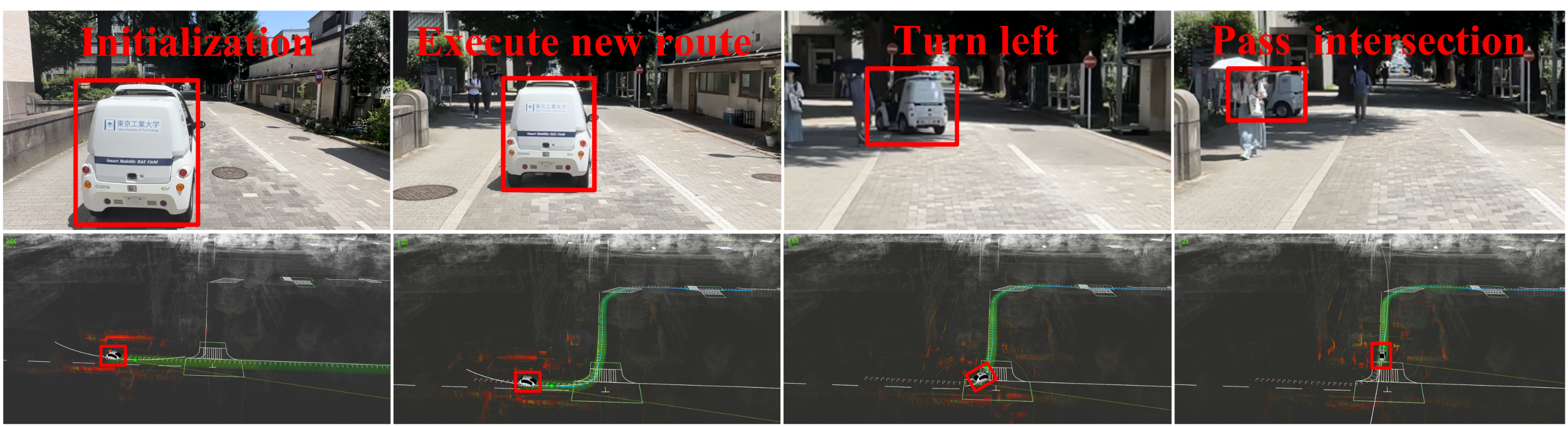} 
\end{minipage}
}
\caption{Autonomous driving operation results: (a) Result \#A: driving on default route, (b) Result \#B: driving on alternative route}
\label{fig: results}
\end{figure*}

The real-time DT modeling results on the cloud plane are illustrated in Fig.~\ref{fig: cloudcases}. As discussed in Section III.A, we upload diverse types of perception data in different application scenarios. In this case, we only transmit object-tracking data to alleviate communication loads. The tracking result has a significantly lower data rate, ranging from $50$ to $300$~Kb/s, depending on the number of detected objects, compared to the raw point cloud data (approximately $46$~Mb/s). Within the LiDAR detection range of the RSU, various entities are identified and visualized using different colored bounding boxes, indicating their categories, positions, and approximate shapes. In the figure, pink boxes represent pedestrians, while blue boxes represent vehicles. The tracking frequency can achieve a rate of $30$~Hz. The cloud server performs fusion and processing of these recognized objects to choose the best global route for the vehicle. We also mark the origin and destination of the autonomous vehicle in this figure, i.e., the vehicle drives from the bottom-left road segment to the bottom-right road segment. In scenario \#A, where no congestion is observed on any road segment, the cloud server selects the straight route indicated by the blue dashed line. In scenario \#B, congestion occurs within the straight-line segment due to the presence of pedestrians and vehicles. In such cases, the route planner selects the path represented by the yellow dashed line, which has a longer distance but effectively helps the autonomous vehicle avoid heavily congested road segments, enabling an efficient and smooth driving process.

The results of the autonomous driving operation, from real-world images and Autoware visualization tool Rviz, are shown in Fig.~\ref{fig: results}. During the initialization phase, the vehicle follows the default straight route in the road network, so in both results, the vehicle initially tracks the straight trajectory. When the ego vehicle arrives at the predetermined threshold distance $D_{\textrm{thre}}$ from the intersection area, it will send a route planning request to the cloud. Result \#A corresponds to scenario \#A, where the vehicle downloads the same default path from the cloud and successfully traverses the intersections in a straight manner. Result \#B corresponds to scenario \#B, where congestion is observed on the straight-line segment. In this case, the vehicle downloads the alternative path from the cloud. The vehicle makes a left turn and proceeds along the alternative route.

In our experiment setting, the threshold distance from the ego vehicle to the intersection is determined with the consideration of safety and comfort issues. The ego vehicle should have enough time to process the planned route before reaching the intersection. Given the presence of pedestrians crossing at intersections, it is necessary for the vehicle to comfortably decelerate and stop before reaching the intersection. In our system, the ego vehicle's free-flow speed $v_{\textrm{f}}$ is set to $15$~km/h. Braking can be classified into emergency and comfortable types. In the presence of unexpected objects, the majority of emergency braking deceleration is more than $4.5$~m/s\textsuperscript{2} \cite{b20}. According to Institute of Transportation Engineers (ITE) recommendations, a comfortable deceleration rate $a_{\textrm{comfy}}$ should be less than $10$ ft/s\textsuperscript{2}, equivalent to $3.048$~m/s\textsuperscript{2} \cite{b21}. Thus the threshold distance is determined as $D_{\textrm{thre}}=0.039 \times v_{\textrm{f}}^2/a_{\textrm{comfy}} = 2.9$~m.

\subsection{Evaluation}

The proposed platform is evaluated from \emph{reliability and latency}. In DT modeling progress, congestion caused by heavy traffic flow is not a typical concern in the campus environment. Instead, congestion may arise during specific periods due to high pedestrian volumes or gatherings. In our demonstration, the frequency of changes in the path selection does not exceed 0.92 changes per minute. Hence, real-time route planning decisions exhibit low sensitivity to time variations. On the other hand, reliability, i.e., the packet delivery rate (PDR) in communication from RSUs to the cloud server, is critical as severe packet loss of tracking results could lead to incorrect planning decisions. If only the tracking results are uploaded from the RSUs, the PDR approaches $100\%$. When simultaneously uploading other levels of data, such as raw LiDAR point clouds, the PDR is around $99.53\%$, which meets the reliability requirement for SSMS, i.e., higher than $95\%$ \cite{b22}.

\begin{figure}[t]
    \centerline{\includegraphics[width=0.5\textwidth]{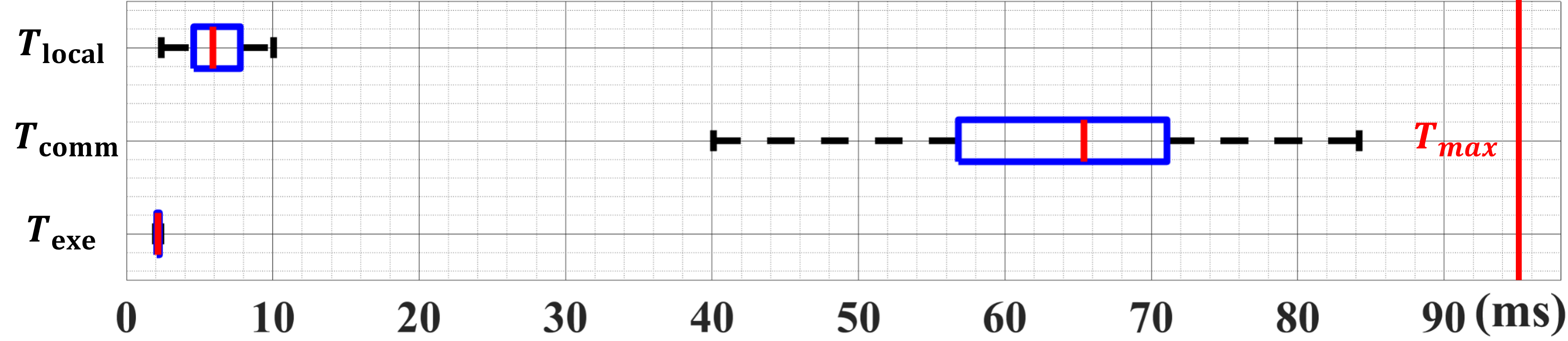}}
    \caption{Latency in route planning service.}
    \label{fig: eval}
\end{figure}

Fig.~\ref{fig: eval} shows the latency in the route request process, including delays caused by localization and position check $T_{\textrm{local}}$, path execution $T_{\textrm{exe}}$ on Autoware PC, and communication $T_{\textrm{comm}}$ with cloud server. The maximum total latency $T_{\textrm{max}}$ is $96.61$~ms, which is $3.39\%$ below the threshold of the max E2E latency requirements for information sharing (less than $100$~ms) proposed by 3GPP \cite{b22}. Assuming the autonomous vehicle maintains the constant free-flow speed during this period, it is still approximately $2.4$~m away from the intersection. This distance provides sufficient room for the vehicle to smoothly maneuver through the intersection without deceleration or, in the presence of pedestrians, to appropriately decelerate ($3.62$~m/s\textsuperscript{2} on average, smaller than the emergency braking deceleration $4.5$~m/s\textsuperscript{2} \cite{b20}) and stop in front of the intersection.

\section{Conclusion}

In this paper, we have designed a smart mobility DT for autonomous driving. Our system utilizes RSUs to capture real-world traffic information, which is processed in the cloud to create a real-time DT model, enabling route planning services for the autonomous vehicle. We have implemented and demonstrated the proposed system in the Tokyo Tech smart mobility field. Test results show that the PDR of DT modeling can reach $99.53\%$ and the latency of route planning service is smaller than $96.61$~ms, which validates the effectiveness of the system in terms of reliability and latency. The latency performance currently leaves room for improvement, indicating that future work should focus on optimizing the communication system.

\section*{Acknowledgment}

This work was supported by the MEXT Doctoral Program for World-leading Innovative \& Smart Education for Super Smart Society (WISE-SSS), JST SPRING (Grant Number JPMJSP2106), NICT-JUNO (Promotion of Commissioned Research for Advanced Communications and Broadcasting Research and Development (\#22404)), and by the U.S. National Science Foundation under JUNO3 Grant CNS-2210254.

\bibliographystyle{IEEEtran.bst}
\bibliography{bibliography.bib}

\vspace{12pt}
\end{document}